\documentclass[aps,twocolumn,prc,floatfix,showpacs,preprintnumbers,amsmath,amssymb,nofootinbib,groupedaddress]{revtex4-2}
\usepackage{bm}
\usepackage[normalem]{ulem}
\usepackage{wasysym}
\usepackage{color}
\usepackage{graphicx}
\usepackage{dcolumn}    
\usepackage{bigstrut}
\usepackage{CJK}
\usepackage[pdfstartview=FitH,
            CJKbookmarks=true,
            bookmarksnumbered=true,
            bookmarksopen=true,
            colorlinks,
            pdfborder=001,
            linkcolor=blue,
            anchorcolor=blue,
            citecolor=blue
            ]{hyperref}

\def\ve#1{{\bm{#1}}}

\def\urm#1{\scriptstyle{\text{\textrm{\textmd{\textup{#1}}}}}}

\let\temp\epsilon
\let\epsilon\varepsilon
\let\varepsilon\temp
\let\temp\relax
\let\temp\phi
\let\phi\varphi
\let\varphi\temp
\let\temp\relax
\begin{document}
%
\title{Isospin symmetry breaking in the charge radius difference of mirror nuclei}

\author{Tomoya Naito${}^{1,2}$, Xavier Roca-Maza${}^{3}$, Gianluca Col\`{o}${}^{3}$, Haozhao Liang${}^{1,2}$, Hiroyuki Sagawa${}^{2,4}$}

\affiliation{
  ${}^{1}$Department of Physics, Graduate School of Science, The University of Tokyo, Tokyo 113-0033, Japan\\
  ${}^{2}$RIKEN Nishina Center, Wako 351-0198, Japan\\
  ${}^{3}$Dipartimento di Fisica, Universit\`{a} degli Studi di Milano, and INFN, Via Celoria 16, 20133 Milano, Italy\\
  ${}^{4}$Center for Mathematics and Physics, University of Aizu,  Aizu-Wakamatsu 965-8560, Japan
}

\date{\today} 

\begin{abstract}
  Isospin symmetry breaking (ISB) effects in the charge radius difference $\Delta R_{\rm ch}$ of mirror nuclei are studied using the test example of ${}^{48}$Ca and ${}^{48}$Ni. This choice allows for a transparent study of ISB contributions since paring and deformation effects, commonly required for the study of mirror nuclei, can be neglected in this specific pair. The connection of $\Delta R_{\rm ch}$ with the nuclear Equation of State and the effect of ISB on such a relation are discussed according to an Energy Density Functional approach. We find that nuclear ISB effects may shift the estimated value for the symmetry energy slope parameter $L$ by more than 10 MeV while Coulomb corrections can be neglected. ISB effects on the ground-state energy and charge radii in mirror nuclei have been recently predicted by {\it ab initio} calculations to be relatively small, pointing to a negligible effect for the extraction of information on the nuclear EoS. These contrasting results call for a dedicated theoretical effort to solve this overarching problem that impacts not only the neutron skin thickness or the difference in mass and charge radii of mirror nuclei but also other observables such as the Isobaric Analog State energy.
\end{abstract}


\maketitle 


There exist important experimental efforts to measure, far from the stability valley, one of the most basic properties of the atomic nucleus: {\it the charge radius} \cite{campbell2016, suda2017}. The study of the charge radius is very much appealing since its determination is free from most nuclear physics uncertainties coming from the strong interaction: the electric charge of a nucleus is determined via electromagnetic probes. Exotic phenomena expected to be observed in those nuclei may help in understanding nuclear structure under extreme conditions of isospin asymmetry \cite{garcia-ruiz2016, gorges2019,chen2019}. At the moment relative isotopic changes in the charge radius can be measured via laser spectroscopy \cite{campbell2016} while two projects, SCRIT \cite{tsukada2017} and ELISe \cite{elise}, aim at measuring the absolute values.

A newly proposed method to explore and estimate the density dependence of the nuclear Equation of State (EoS) of isospin asymmetric matter requires the measurement of the charge radii in mirror mass nuclei \cite{brown2017}. Specifically, it has been shown in \cite{brown2017} on the basis of different Energy Density Functionals (EDFs) that the difference in the charge root-mean-square radius of the nucleus with $N$ neutrons and $Z$ protons, $R_{\rm ch}^{(N,Z)}$, with respect to its mirror nucleus, that has the number of neutrons and protons interchanged, $\Delta R_{\rm ch}\equiv R_{\rm ch}^{(N,Z)}-R_{\rm ch}^{(Z,N)}$, is connected to the symmetry energy slope parameter $L$. The latter is {\it strictly} proportional to the pressure in infinite neutron matter at nuclear saturation density (about 0.16 fm$^{-3}$) only if Isospin Symmetry Breaking (ISB) effects are neglected. A similar relation was already found by the same author \cite{brown2000}: between the neutron skin thickness $\Delta R_{np}\equiv R_n^{(N,Z)}-R_p^{(N,Z)}$ of the nucleus with $N$ neutrons and $Z$ protons and the $L$ parameter (cf. also \cite{roca-maza2011}). On this regard, in Ref.~\cite{roca-maza2018a}, it was shown that the estimation of the neutron skin thickness in ${}^{208}$Pb via the measurement of the excitation energy of the Isobaric Analog State (IAS) is clearly dependent on ISB effects: {\it the larger the neutron skin in ${}^{208}$Pb, the larger ISB effects must be to reproduce the experimental data on IAS}.
If electromagnetic and nuclear ISB contributions are neglected, the excitation energy of the IAS in the nucleus ($N$, $Z$) is zero and $\Delta R_{\rm ch}=-\Delta R_{np}$ being both of the order of a fraction of one fm or less. Hence, the effect of ISB terms other than those originated from the Coulomb interaction may become relevant. 

Due to the impact of the nuclear EoS and, in particular, of the $L$ parameter on areas as diverse as nuclear physics and nuclear astrophysics \cite{baldo2016,oertel2017,roca-maza2018b}, the study of $\Delta R_{np}$ have fostered different experiments and a number of theoretical investigations along the years (see, e.g., \cite{tsang2012,horowitz2014,fattoyev2018,prex,prex2,sotani2022}). Similar situation is starting to take place regarding the measurement of $\Delta R_{\rm ch}$. As an example, the mirror pairs ${}^{36}$Ca-${}^{36}$S and ${}^{38}$Ca-${}^{38}$Ar have been analyzed in \cite{brown2020} and ${}^{54}$Ni-${}^{54}$Fe in \cite{pineda2021} on the basis of nuclear EDFs. Those studies neglect ISB effects as well as pairing correlations, and also deformation is not treated microscopically. Actually, in a recent study based on quantified EDFs \cite{reinhard2022}, $\Delta R_{\rm ch}$ has been shown to be influenced by pairing correlations in the presence of low-lying proton continuum --in the proton-rich partner-- and the authors concluded that, considering the large theoretical uncertainties, precise data on mirror charge radii cannot provide a stringent constraint on $L$.

In the present work, we concentrate on a different source of systematic uncertainty with respect to those discussed in Ref.~\cite{reinhard2022}. It will be shown that nuclear ISB contributions cannot be neglected in the study of the difference of charge radii in mirror nuclei if analysed on the basis of current EDFs. For that, the specific example ${}^{48}$Ca and ${}^{48}$Ni mirror nuclei is used. This choice avoids the effects of pairing correlations and deformation on $\Delta R_{\rm ch}$ since both nuclei are predicted to be doubly-magic in our calculations.

{\it Nuclear model calculations.} There exist different theoretical approaches to the description of the charge radii in nuclei. One of the most successful nowadays is based on nuclear Density Functional Theory \cite{skyrme1,colo2018,schunck2019}. State-of-the-art nuclear EDFs are known to provide predictions of experimentally known charge radii along the whole nuclear chart which are at the level of 0.02-0.03 fm of average accuracy \cite{goriely2013,afanasjev2016,roca-maza2018b,reinhard2021}. This accuracy is not reached by other approaches available in the literature, hence the suitability of this theoretical framework.

For the present study, we use the SAMi-J family of EDFs \cite{roca-maza13}. We will see at the end that our main conclusions are independent of this choice. Except for the Coulomb interaction, these Skyrme-type functionals are isospin symmetric --adopting the most standard form found in the literature \cite{skyrme1}-- and have been calibrated using the fitting protocol of the SAMi EDF \cite{roca-maza12b}.
The SAMi-J family has been produced by exploring the optimal parameterization around the minimum (SAMi) in terms of the variation of two parameters that characterize the EoS of asymmetric nuclear matter: the symmetry energy at nuclear saturation density $J$ and the previously introduced $L$ parameter. We will show results for EDFs with $J$ ($L$) ranging values from 27 (30) MeV to 35 (115) MeV. This range spans the accepted values for these quantities \cite{roca-maza2018b}. On top of SAMi-J EDFs, we will first implement and present the effects on $\Delta R_{\rm ch}$ originated from the Coulomb interaction in a similar fashion to what has been done in Ref.~\cite{roca-maza2018a}. As we shall discuss later, Coulomb effects are not accounted for in the nuclear EoS, hence, the $L$ parameter is by construction insensitive to the Coulomb interaction. Subsequently, we will explore and present the effects of ISB contributions coming from the nuclear strong interaction that will affect both the $\Delta R_{\rm ch}$ and the properties of the nuclear EoS. Finally, we will present all effects combined. 

{\it Coulomb interaction.} In standard EDFs it is customary to account for the Coulomb interaction in the atomic nucleus by using the Hartree approximation for the direct term and the Slater approximation for the exchange term. This is named in the literature Local Density Approximation (LDA) to the Coulomb part of the EDF. While this is reasonable for the description of different observables, it may become inaccurate for the present case. Hence, we will also show results by adopting the Generalized Gradient Approximation (GGA) that accurately reproduces the exact Fock results \cite{naito2018, naito2019}.

In addition to that, since $\Delta R_{\rm ch}$ is a small quantity, other corrections beyond the GGA could be relevant. Those corrections are the electromagnetic finite size effects of the nucleons and leading order Quantum Electrodynamic corrections --vacuum polarization-- to the Coulomb interaction. In Refs.~\cite{roca-maza2018a, naito2020} these corrections are described in detail.

\begin{figure}[t!]
\includegraphics[width=\linewidth,clip=true]{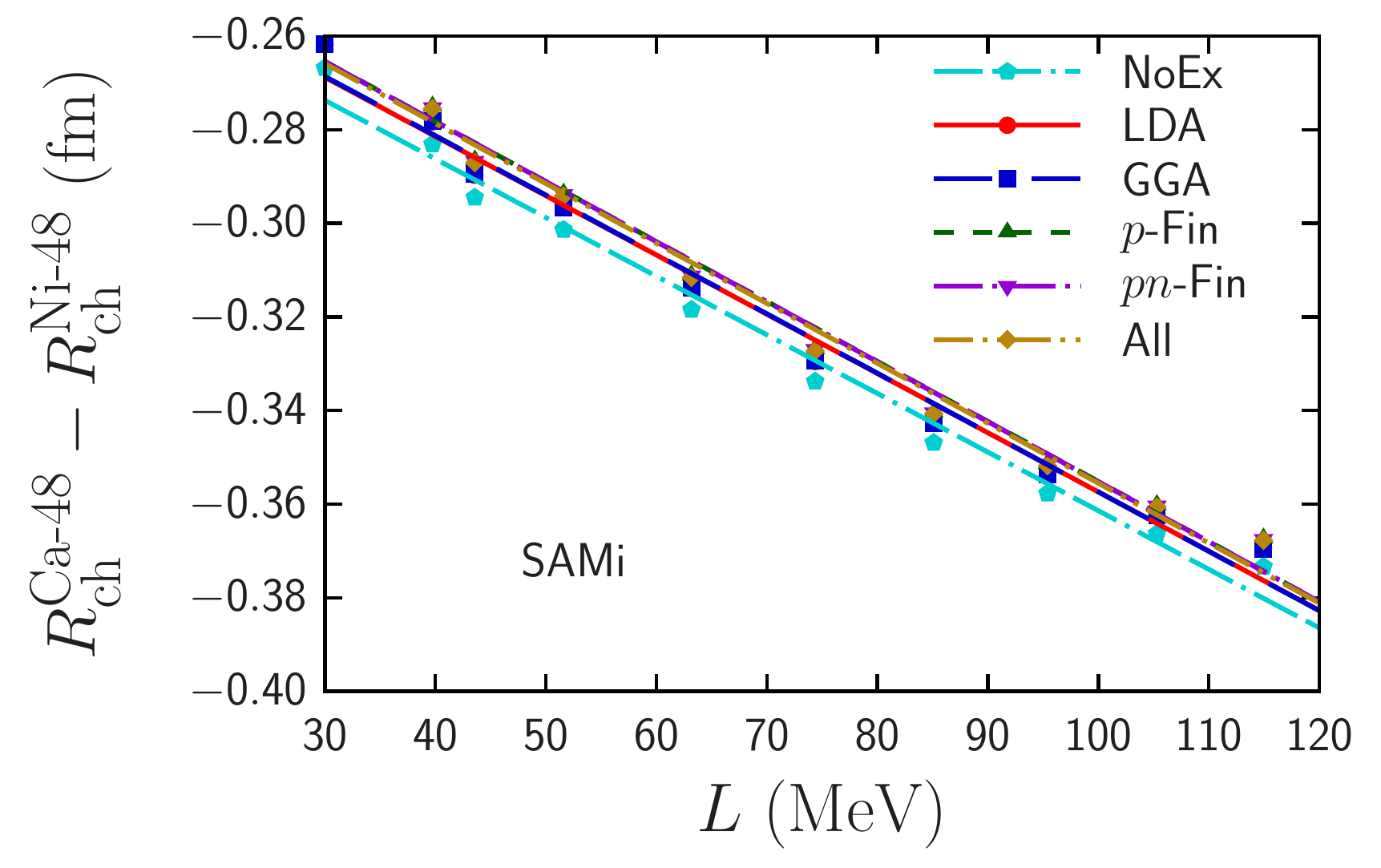}
\caption{SAMi-J predictions for the charge radius difference between ${}^{48}$Ca and ${}^{48}$Ni as a function of the $L$ parameter. Different approximations to the Coulomb interaction has been adopted (see text for details).}
\label{fig1}
\end{figure}

In Fig.~\ref{fig1} we show that the different Coulomb corrections and approximations taken into account barely modify the values of the difference between the charge radii in ${}^{48}$Ca and ${}^{48}$Ni. The correlation between these values and the $L$ parameter as proposed in \cite{brown2017} is, therefore, also unaffected. For guidance, average linear trends are given by lines. The results shown have been produced with the SAMi-J EDFs, including the unconstrained one: SAMi. In the figure, ``NoEx'' stands for calculations neglecting the Coulomb exchange term of the functional. Neglecting 
exchange terms has been postulated as a phenomenologic recipe to compensate nuclear ISB effects on the masses of mirror nuclei \cite{brown1998,brown2000,goriely2008}. Then, adding to the latter, improving approximations and corrections are implemented. First the Coulomb exchange is taken into account in the LDA and subsequently in the GGA. Two different electromagnetic finite size effects have been implemented in the calculation of the charge radii and Coulomb potential: the first ``$p$-Fin'' takes into account only the electric from factor of the proton while ``$pn$-Fin'' adds to the previous one also the neutron electric form factor effects. Finally, labeled as ``All'', leading order Quantum Electrodynamic correction to the Coulomb interaction and magnetic form factors effects have been added.    

{\it Nuclear isospin symmetry breaking terms.} The isospin symmetry breaking of the nuclear interaction can be divided into two parts; the charge symmetry breaking (CSB) interaction $V_{\rm CSB} \equiv V_{nn}-V_{pp}$ and the charge independent breaking (CIB) interaction $V_{\rm CIB} \equiv (V_{nn}+V_{pp})/2 -V_{pn}$, where $V_{pp}$, $V_{nn}$, and $V_{pn}$ denote the proton-proton, neutron-neutron, and proton-neutron nuclear interactions, respectively. Origins of CSB are mainly due to proton-neutron mass difference as well as $\pi^0-\eta$ and $\rho^0-\omega$ mixings in meson exchange processes while that of CIB is the mass difference of neutral and charged pions \cite{miller1990}.

The form of the CSB and CIB interactions used in this work are \cite{sagawa1995,roca-maza2018a}
\begin{subequations}
  \label{eq1}
  \begin{align}
    v_{\urm{CSB}} \left( \ve{r}_1, \ve{r}_2 \right)
    & =
      \frac{\tau_{1z} + \tau_{2z}}{4}
      s_0
      \left( 1 + y_0 P_{\sigma} \right)
      \delta \left( \ve{r}_1 - \ve{r}_2 \right), 
      \label{eq1a} \\
    v_{\urm{CIB}} \left( \ve{r}_1, \ve{r}_2 \right)
    & =
      \frac{\tau_{1z} \tau_{2z}}{2}
      u_0
      \left( 1 + z_0 P_{\sigma} \right)
      \delta \left( \ve{r}_1 - \ve{r}_2 \right), 
      \label{eq1b}
  \end{align}
\end{subequations}
respectively, where $P_{\sigma} = \left( 1 + \ve{\sigma}_1 \cdot \ve{\sigma}_2 \right) / 2 $ is the spin-exchange operator, $\ve{\sigma}$ the Pauli matrices in spin space and $ \tau_{iz} $ is the $z$ component of the Pauli matrices in isospin space. These forms are two of the simplest possible yet realistic enough to describe the IAS in different nuclei \cite{roca-maza2018a}. Other functional forms of these interactions have been discussed in Refs.~\cite{doba2019,doba2021}.

{\it Equation of State.} The nuclear EoS at zero temperature corresponds to the energy per particle ($e\equiv E/A$) of an infinite sea of neutrons and protons at a fixed constant density where the Coulomb interaction is not taken into account for obvious reasons.

According to the definition above, the Hartree-Fock CSB and CIB contributions to the energy per particle can be derived in terms of the total density $\rho=\rho_n+\rho_p$ and the relative difference $\beta=(\rho_n-\rho_p)/\rho$ between the neutron ($\rho_n$) and proton ($\rho_p$) density distributions as~\cite{naito2020}
\begin{subequations}
  \label{eq2}
  \begin{align}
    e_{\urm{CSB}} \left( \rho, \beta \right)
    & =
      \frac{s_0 \left( 1 - y_0 \right)}{8}\rho\beta \ ,
      \label{eq2a} \\
    e_{\urm{CIB}} \left( \rho, \beta \right)
    & =
      -\frac{1}{16}u_0(1+2z_0)\rho + \frac{3}{16}u_0\rho\beta^2 \ , 
      \label{eq2b}
  \end{align}
\end{subequations}
respectively. 

In nuclear physics, it is customary to expand the energy per particle around $\beta\rightarrow 0$ up to $\mathcal{O}[\beta^2]$. This approximation has been shown to be reasonable even for large values of $\beta$, provided one remains at densities below two times saturation density \cite{vidana2009}. This is actually the situation here. Thus, without loosing generality for our current purposes, the EoS can be written as,
\begin{equation}
e(\rho,\beta) = e_0(\rho)+e_1(\rho)\beta+e_2(\rho)\beta^2+\mathcal{O}[\beta^3] \ ,
\label{eq3}
\end{equation}
where $e_0(\rho)$ is the EoS of symmetric nuclear matter ($\beta=0$) and gets contributions from CIB in Eq.~(\ref{eq2b}), $e_1(\rho)$ is a contribution originated by the CSB interaction in Eq.~(\ref{eq2a}) and $e_2(\rho)$ is the symmetry energy as commonly defined in the literature. The latter term gets also contributions from CIB in Eq.~(\ref{eq2b}). In the standard definition of the EoS that assumes the isospin symmetry (IS): $e_1(\rho)$ would be zero and CIB contributions to $e_0(\rho)$ and $e_2(\rho)$ would also be zero. That specific case is when the $J$ and $L$ parameters are customarily defined by expanding $e_2(\rho)$ around saturation density ($\rho_0$) as,
\begin{equation}
e_2^{\rm IS}(\rho)= J + L\epsilon + \mathcal{O}[\epsilon^2] \ ,
\label{eq4}
\end{equation}
where $\epsilon\equiv (\rho-\rho_0)/3\rho_0$. 

{\it Neutron matter pressure at saturation density.} Assuming isospin symmetry, the contribution to the total pressure of neutron matter ($\beta=1$) at saturation would be proportional to $L$,
\begin{equation}
P^{\rm IS}(\rho_0,\beta=1) = \rho^2\frac{\partial e(\rho,\beta)}{\partial \rho}\Bigg\vert_{\rho=\rho_0} = \frac{1}{3}\rho_0 L\ , 
\label{eq5}
\end{equation}
but this is not the case if nuclear ISB terms are included,
\begin{equation}
P(\rho_0,\beta=1) = \rho^2\frac{\partial e(\rho,\beta)}{\partial \rho}\Bigg\vert_{\rho=\rho_0} = \frac{1}{3}\rho_0 (L_{\rm CIB} + L_{\rm CSB} + L)\ . 
\label{eq6}
\end{equation}
In the last equation,
\begin{subequations}
  \label{eq7}
  \begin{align}
  L_{\rm CSB} \equiv& \frac{3}{8}s_0\rho_0(1-y_0) \ ,
      \label{eq7a} \\
  L_{\rm CIB} \equiv& \frac{9}{16}u_0\rho_0 \ .    
  \end{align}
\end{subequations}
In Ref.~\cite{roca-maza2018a}, the values of the ISB parameters built on top of the SAMi EDF (SAMi-ISB) have been found in a semi-phenomenological way. This functional reproduce CIB contributions in symmetric nuclear matter as calculated in \cite{muther1999} while CSB has been fixed to reproduce the IAS in ${}^{208}$Pb. The values are $s_0=-26.3\pm0.7$ MeV fm$^3$ and $u_0=25.8\pm 0.4$ MeV fm$^3$  with $y_0$ and $z_0$ fixed to $-1$. These values would imply a change in the neutron pressure at saturation in Eq.~(\ref{eq5}) --that has typical values around 3 MeV fm${}^{-3}$-- of about $-0.17$ MeV fm${}^{-3}$ due to CSB and +0.12 MeV fm${}^{-3}$ due to CIB, hence, cancelling to a large extent. This partial cancellation must be investigated in more detail since the CSB and CIB effects induce fundamental differences between $P(\rho_0,\beta=1)$ and $L$.
It is worth noting that in Ref.~\cite{BACZYK2018}, where essentially the same CSB interaction was assumed, a value of $s_0$  that could range from $-8.4$ to $-18.6$ MeV fm$^3$ was reported (cf. Table 1 in Ref.~\cite{BACZYK2018} where $s_0=2t_0^{\rm III}$).       

In Table I of Ref.~\cite{novario2021}, {\it ab initio} Coupled Cluster calculations based on the $\Delta$N$^2$LO$_{\rm GO}$(394) interaction are shown for $\Delta R_{\rm ch}$ and binding energy difference ($\Delta B$) in ${}^{48}$Ca-${}^{48}$Ni. The reported value for $\Delta B$ is 66.29$\pm$0.72 MeV including all effects and 0.72$\pm$0.01 MeV by neglecting Coulomb effects. The $\Delta B$ estimated value from the Atomic Mass Evaluation (AME2020) \cite{ame2020} is 68.67 MeV. The EDF proposed in Ref.~\cite{roca-maza2018a}, SAMi-ISB, predicts $\Delta B=68.72$ MeV, with an effect due to ISB of 7.88 MeV. Since $\Delta B$ is only due to nuclear ISB when Coulomb is neglected, it is evident that the latter EDF results are not in agreement with {\it ab initio} calculations in \cite{novario2021}. In more detail, by following the strategy proposed in \cite{naito2021}, we have determined an effective $s_0$ value of $-2$ MeV fm$^3$ associated with the {\it ab initio} results for $\Delta B$ in ${}^{48}$Ca-${}^{48}$Ni given in \cite{novario2021}. This value is one order of magnitude smaller than those from Refs.~\cite{roca-maza2018a,BACZYK2018}.

Variational Monte Carlo calculations \cite{wiringa} based on AV18 \cite{wiringa1995} for $\Delta B$ in ${}^{10}$Be-${}^{10}$C would predict an effective $s_0$ of about $-3$ MeV fm$^3$ when using again the method proposed in \cite{naito2021}. Thus, they give a similar value to the one we have just discussed~\cite{novario2021}. Regarding $\Delta R_{\rm ch}$ in ${}^{48}$Ca-${}^{48}$Ni, the prediction in Ref.~\cite{novario2021} is of 0.238$\pm$0.038 fm for the full calculation while it is 0.261$\pm$0.035 fm for the calculation neglecting Coulomb effects. SAMi-ISB predict instead 0.33 fm for the full calculation and 0.28 fm for the calculation neglecting Coulomb effects. These results reflect an opposite trend in $\Delta R_{\rm ch}$. Naively, the Coulomb potential ($Ze^2/r$) would be expected to increase the charge radius, more in absolute value as larger the $Z$ of the nucleus. Hence, $\Delta R_{\rm ch}$ would be expected to be larger when Coulomb is included. These contrasting results call for a dedicated theoretical effort since observables as relevant as the excitation energy of the Isobaric Analog State, the neutron skin thickness, or the difference in mass and charge radii of mirror nuclei would be affected.    

\begin{figure}[t!]
\includegraphics[width=\linewidth,clip=true]{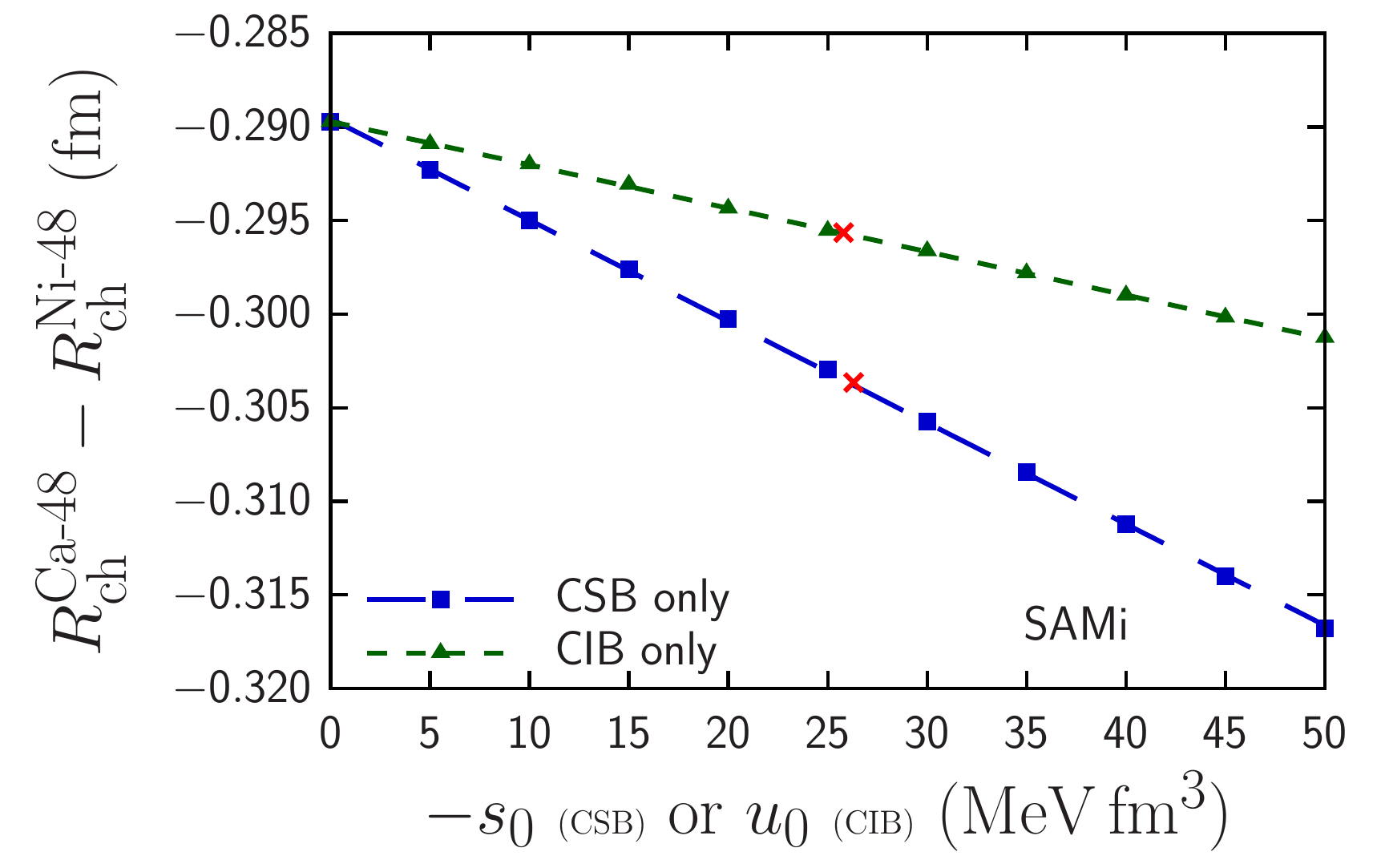}
\caption{SAMi EDF predictions for the charge radius difference between ${}^{48}$Ca and ${}^{48}$Ni as a function of the CSB $-s_0$ (blue squares) and CIB $u_0$ (green triangles) parameters. As a reference, the predictions in Ref.~\cite{roca-maza2018a} for those parameters are given by red crosses. The values of $y_0$ and $z_0$ have been fixed to -1.}
\label{fig2}
\end{figure}

In Fig.~\ref{fig2}, we show the predictions of the SAMi EDF for the charge radius difference between ${}^{48}$Ca and ${}^{48}$Ni. Nuclear ISB contributions have been included perturbatively. Values of $s_0$ and $u_0$ ($y_0=z_0=-1$) have been changed from 0 to 50 MeV fm$^3$ and the effects shown separately in the figure. As a reference, the values suggested in Ref.~\cite{roca-maza2018a} are given by red crosses. It is important to note that both CSB and CIB are {\it coherent} and tend to make $\Delta R_{\rm ch}$ larger in absolute value; by about 6\% if $-s_0\approx u_0\approx 25$ MeV fm$^3$.  

These trends can be understood from the definition of the ISB interactions in Eq.~(\ref{eq2}). The CSB average potential felt by protons is proportional to the proton density and $-s_0$. This means that for $s_0<0$ protons feel a CSB repulsive potential that grows linearly with $\sim Z$ and, thus, the charge radii tends to increase with $Z$ as well. The CIB average potential felt by protons is proportional to $u_0$ and to the difference $\sim Z-N/2$. For $u_0>0$ protons feel a repulsive CIB average potential, however, the charge radii do not grow as fast as for CSB case.

\begin{figure}[t!]
\includegraphics[width=\linewidth,clip=true]{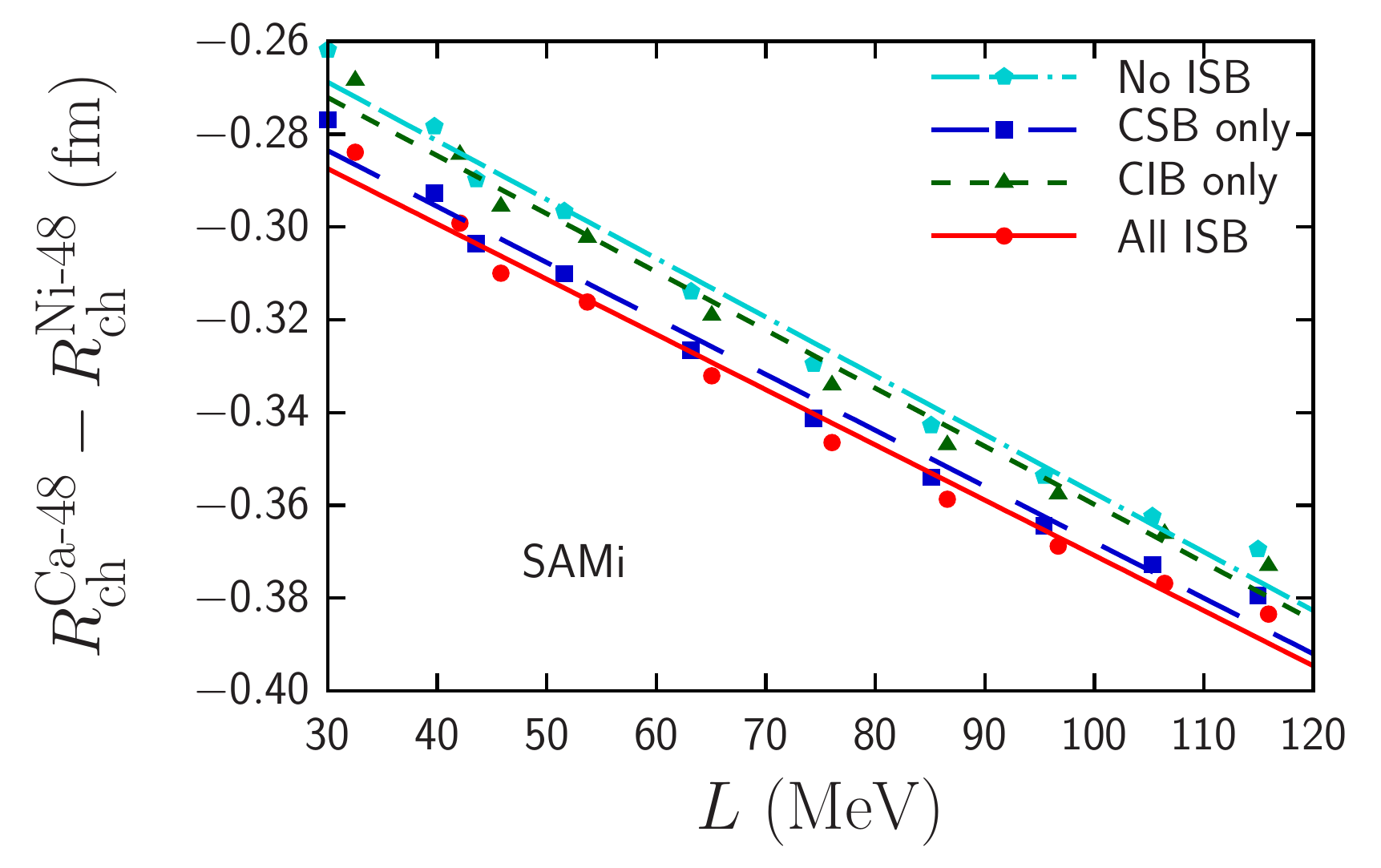}
\caption{$\Delta R_{\rm ch}$ between ${}^{48}$Ca and ${}^{48}$Ni as a function of the $L$ parameter is shown as predicted by the SAMi-J family of EDFs. Different contributions are shown. Original SAMi-J given by circles and labeled by ``No ISB''; SAMi-J plus CSB in Eq.~(\ref{eq2a}) given as squares are labeled by ``CSB only''; SAMi-J plus CIB in Eq.~(\ref{eq2b}) given as triangles are labeled by ``CIB only''; and SAMi-J with all corrections given as circles is labeled as ``All ISB''. The latter includes Coulomb within the LDA.}   
\label{fig3}
\end{figure}

In Fig.~\ref{fig3}, predictions of the SAMi-J family are given for $\Delta R_{\rm ch}$ between ${}^{48}$Ca and ${}^{48}$Ni as a function of the $L$ parameter. In this figure, calculations include Coulomb within the LDA. It is seen that the main correction to the charge radius difference between the mirror nuclei ${}^{48}$Ca and ${}^{48}$Ni is due to CSB while CIB remains small. The average horizontal separation between the results is slightly larger than 11 MeV (9 MeV comming form CSB) and rather constant for the large range of values of $L$ displayed in that figure.

{\it Conclusions.} The effects discussed here are based on the ISB interactions in Eq.~(\ref{eq1}) but are independent of the EDF used as a basis. This is shown by the constant shift in Fig.~\ref{fig3}. Hence, the different trends can be regarded as general provided the ISB interactions in Eq.~(\ref{eq1}) are realistic enough for the current purpose. Our results indicate that one must expect a correction: {\it given $\Delta R_{\rm ch}$, the estimated value of $L$ must be decreased whenever ISB are neglected}. The correction due to nuclear ISB based on our calculations of the mirror pair ${}^{48}$Ca-${}^{48}$Ni would correspond to a shift of about 10 MeV and, even if not large, it needs to be incorporated --among other corrections \cite{reinhard2022}-- by a sound study of $\Delta R_{\rm ch}$. For the same reasons, the neutron skin thickness $\Delta R_{np}$ in a neutron-rich nucleus is also affected by ISB effects.

The situation is different in {\it ab initio} calculations \cite{novario2021,wiringa} where the estimated effective value of $s_0$ seems to be one order of magnitude smaller than those inferred from current EDFs \cite{roca-maza2018a,BACZYK2018}. These contrasting results call for a dedicated theoretical effort to solve this overarching problem that impacts the energy of the Isobaric Analog State, the neutron skin thickness, or the difference in mass and charge radii of mirror nuclei as well as it may impact astrophysical observables such as the mass, radius, and tidal deformability of a neutron star \cite{selva2021}. 

{\it Acknowledgements.} The authors thank S. Gandolfi and A. Ekstr\"om for useful discussions as well as R.~B. Wirnga for providing us with Variational Monte Carlo calculations for $\Delta B$ in ${}^{10}$Be-${}^{10}$C. T.N. and H.L. thank the RIKEN iTHEMS program and the RIKEN Pioneering Project: Evolution of Matter in the Universe. T.N. acknowledges the JSPS Grant-in-Aid for JSPS Fellows under Grant No. 19J20543. H.L. acknowledges the JSPS Grant-in-Aid for Early-Career Scientists under Grant No. 18K13549 and the Grant-in-Aid for Scientific Research (S) under Grant No. 20H05648. H.S. acknowledges the Grant-in-Aid for Scientific Research (C) under Grant No. 19K03858. The numerical calculations were performed on cluster computers at the RIKEN iTHEMS program. 

\bibliography{bibliography.bib}

\end{document}